\documentclass[sigconf]{acmart}

\AtBeginDocument{%
  }

\copyrightyear{2025}
\acmYear{2025}
\setcopyright{cc}
\setcctype{by}
\acmConference[CHI '25]{CHI Conference on Human Factors in Computing Systems}{April 26-May 1, 2025}{Yokohama, Japan}
\acmBooktitle{CHI Conference on Human Factors in Computing Systems (CHI '25), April 26-May 1, 2025, Yokohama, Japan}\acmDOI{10.1145/3706598.3713931}
\acmISBN{979-8-4007-1394-1/25/04}

\acmConference[CHI '25]{CHI Conference on Human Factors in Computing Systems}{April 26-May 1, 2025}{Yokohama, Japan}
\acmISBN{979-8-4007-1394-1/25/04}

\usepackage{xcolor}
\usepackage{makecell}

\usepackage{graphicx}

\begin{document}

\renewcommand{\arraystretch}{1.5} 

\title[Secure Software Demonstration for Software SMEs in a B2B Model]{No Silver Bullet: Towards Demonstrating Secure Software Development for Danish Small and Medium Enterprises in a Business-to-Business Model}

\author{Raha Asadi}
\email{asad@itu.dk}
\orcid{0009-0005-1109-9696}
\affiliation{%
  \institution{IT University of Copenhagen}
  \city{Copenhagen}
  \country{Denmark}
}

\author{Bodil Biering}
\email{bodil@cyberjuice.io}
\orcid{0009-0000-0830-2968}
\affiliation{%
  \institution{Cyberjuice}
  \city{Copenhagen}
  \country{Denmark}
}

\author{Vincent van Dijk}
\email{vincent@securityscientist.net}
\orcid{0009-0003-4600-6440}
\affiliation{%
  \institution{Security Scientist}
  \city{Copenhagen}
  \country{Denmark}
}

\author{Oksana Kulyk}
\email{okku@itu.dk}
\orcid{0000-0003-4218-1658}
\affiliation{%
  \institution{IT University of Copenhagen}
  \city{Copenhagen}
  \country{Denmark}
}

\author{Elda Paja}
\email{elpa@itu.dk}
\orcid{0000-0002-8346-2467}
\affiliation{%
  \institution{IT University of Copenhagen}
  \city{Copenhagen}
  \country{Denmark}
}

\renewcommand{\shortauthors}{Asadi et al.}

\begin{abstract}
Software developing small and medium enterprises (SMEs) play a crucial role as suppliers to larger corporations and public administration. It is therefore necessary for them to be able to demonstrate that their products meet certain security criteria, both to gain trust of their customers and to comply to standards that demand such a demonstration. In this study we have investigated ways for SMEs to demonstrate their security when operating in a business-to-business model, conducting semi-structured interviews ($N=16$) with practitioners from different SMEs in Denmark and validating our findings in a follow-up workshop ($N=6$). Our findings indicate five distinctive security demonstration approaches, namely: \textit{Certifications}, \textit{Reports}, \textit{Questionnaires}, \textit{Interactive Sessions }and \textit{Social Proof}. We discuss the challenges, benefits, and recommendations related to these approaches, concluding that none of them is a one-size-fits all solution and that more research into relative advantages of these approaches and their combinations is needed. 
\end{abstract}

\ccsdesc[500]{Security and privacy~Social aspects of security and privacy}

\keywords{Security; Commerce/Business; Qualitative methods}

\received{12 September 2024}
\received[revised]{10 December 2024}
\received[accepted]{17 January 2025}

\maketitle

\section{Introduction}

Small and medium enterprises (SMEs) comprise the vast majority of economic activities in the European Union~\cite{EU_SME_Website}, as well as the majority of software development companies ~\cite{GAITERO2021110960}. As a result, the cybersecurity of services and products created and provided by these software development companies becomes of utmost importance, especially since they are often part of supply chains for critical infrastructure IT systems~\cite{CFC2019, Pfeifer, Thun2011}. Consequently, clients, including businesses that depend on the products and services offered by said SMEs, must require assurance of the cybersecurity measures in place. 

In addition to this, the recently proposed NIS2 directive requires companies in the critical infrastructure supply chain, or those associated with the day-to-day operations of a country's economy, to ensure their levels of security to external regulatory and oversight entities \cite{nis2_requirements,affected_by_nis2}.

One common way to provide such guarantees is by using certifications according to security standards set by international organizations (such as ISO \cite{ISO}, NIST \cite{NIST} or COBIT \cite{ISACA_COBIT}) or individual governments (such as the Danish D-Seal, "D-Mærket" \cite{dmaerket}); however, previous research shows that certifications are often too costly and complicated to implement for SMEs~\cite{GAITERO2021110960,OzkanandSpruit,Guerra2021TheRO}. Therefore, both SMEs and clients may need to rely on alternative methods of demonstrating security -- that is, methods to provide certainty or assurances to clients that the software and services provided by SMEs meet the required level of security, which are more accessible to SMEs. Although several studies considered alternative methods such as security labels~\cite{RaeandPatel,cavenetal,chenetal} or testing methods~\cite{Thomasetal10.1145/3173574.3173836,Tahaei2019ASO}, to the best of our knowledge, none of these studies aimed to provide a comprehensive overview of how these and other methods are used in practice, in particular, in the context of providing a high level of confidence to clients. This paper, examines the practical methods of \emph{security demonstrations} that are used by SMEs within the business-to-business (B2B) model. We conduct exploratory research and our work considers both assurances with a high degree of rigor, as well as less formal demonstrations that nonetheless serve the goal of reassuring the stakeholders. 

However, prior to investigating the methods SMEs use to demonstrate security, we first assess whether SMEs understand the requirements expected from them, and we identify who is imposing these demands. Our work therefore addresses the following research questions:

\begin{description}
    \item[RQ1] What are the requirements for the security demonstration imposed on SMEs and where do these requirements come from?
    \item[RQ2] How do SMEs demonstrate the security of their software and services?
\end{description}

To answer these questions, we conducted a qualitative study through semi-structured ethnographic interviews with 16 practitioners from 15 Danish SMEs to identify key security demonstration methods employed by SMEs. Based on the data analysis of the interviews, we also conducted a validation workshop to confirm our findings and gather additional feedback that, in combination with the interview results, culminated in our recommendations. We relied on both the interviews and the validation workshop as a foundation for compiling a set of recommendations for industry, academia and regulators. 

Our study therefore makes three contributions: 

\begin{itemize}
    \item We provide examples of requirements for security demonstrations, spanning certifications, security questionnaires, sociotechnical requirements specified in contracts and regulatory requirements among others. We further identify who imposes these requirements, ultimately distinguishing between clients (large vs small) and legislation.
 
    \item We identify five key methods that SMEs use for security demonstrations. We identify both \emph{passive methods} (i.e., the ones that do not require direct client interaction), namely, \textit{certifications}, \textit{tests and reports}, and \textit{social proof}), and \emph{active methods} (i.e., those that involve client participation), namely, \textit{questionnaires} and \textit{interactive sessions}. Note that while our initial inquiry focused on methods for demonstrating the security of the finished \emph{product}, that is, software products and services provided by SMEs, our participants furthermore mentioned methods they use to demonstrate the security of their overall \emph{infrastructure} and \emph{internal processes}. This finding stresses that both types of security are closely intertwined and that security of products cannot be properly ensured without security of infrastructure and processes.
    \item We discuss the identified methods in terms of how they vary across several factors (e.g. costs, reliability, accessibility) and present a set of recommendations for each security demonstration method to enable SMEs to effectively integrate them within their organization and corresponding software solutions. 
\end{itemize}

\section{Background} \label{sec:Background}
To the best of our knowledge, there is little research on the topic of security demonstrations in particular. At the same time, a number of studies have been conducted to investigate issues related to such security demonstrations, namely, (i) specification and implementation of security requirements, (ii) individual demonstration methods, in particular, labeling and certification, and (iii) further security practices in companies. 

\subsection{Specification and Implementation of Security Requirements}
Security requirements come from a number of factors influencing a company's security initiatives. As such, some researchers argue that adhering to security requirements based on certifications stem from both external pressures (e.g. due to the need to meet customer demands or to qualify for participation in public tenders) or an internal desire for improvement~\cite{GAITERO2021110960}. The implementation of robust cybersecurity practices can provide SMEs with a competitive advantage in the market and lead to lucrative contracts \cite{Chidukwani}.

At the same time, research shows that security requirements can often get neglected or underprioritised due to both internal and external pressures, including business-related demands, budget constraints, and other more visible and measurable tasks~\cite{Tahaei2019ASO}. As such, a study by Assal and Chiasson shows that while software developers are primarily self-motivated to consider security in their development processes, they often face significant challenges, including a lack of security knowledge, insufficient tools, and a lack of organizational prioritization of security~\cite{chiassonassal}.

\subsection{Labeling and Certification}
One effective way to document a company's security is by adhering to industry-recognized security standards and frameworks. Compliance with these standards typically results in a certification number or report, provided that the company chooses to apply for an official certification and pays the associated fees,  indicating that the company meets the minimum criteria established by the certifying body \cite{Guerra2021TheRO,GAITERO2021110960,Chidukwani,cavenetal,chenetal}. The International Organisation for Standardisation (ISO) mandates that any certification is valid for three years, after which a certification renewal audit must be carried out \cite{GAITERO2021110960}. At the same time, the extent to which certifications provide a comprehensive picture of a company’s security is a topic of debate. Guerra and Hinde, for instance, examine the role of product safety certifications in the context of commercial off-the-shelf (COTS) devices used in Nuclear Power Plants (NPPs). They argue that compliance certifications alone are insufficient for full assurance and advocate for a more comprehensive approach that includes behavior assurance, vulnerability assessments, and standards compliance \cite{Guerra2021TheRO}. Similarly, Chidukwani et al. argue that relying on a single framework may hinder companies in achieving their overall business objectives and meeting compliance requirements. They suggest that adopting a hybrid framework would be more effective \cite{Chidukwani}. Certifications furthermore can be challenging to obtain for SMEs due to their limited financial and human resources, as well as a lack of technical expertise in security management \cite{GAITERO2021110960,Chidukwani}. 

There has been proposed alternative strategies for assisting SMEs in tackling and demonstrating security focus on labeling schemes. For example, Rae and Patel propose a consumer-friendly cybersecurity rating system tailored for SMEs in the UK, inspired by the UK's Food Standards Agency’s Food Hygiene Rating Scheme (FHRS). This scheme assigns a security rating from 1 (worst) to 5 (best), based on audits of the SME's security practices and technical systems, and displays the rating at business entrances, payment areas, and online \cite{RaeandPatel}. In the UK, the Cyber Essentials badge targets SMEs by defining five key technical requirements to protect against cyberattacks \cite{govuk_cyber_essentials}. Denmark’s IT Security Badge, D-Mærket \cite{dmaerket}, offers similar guidance. Similar labeling strategies have been discussed for IoT products in B2C models, which could complement our study \cite{chenetal, cavenetal}. These examples of solutions suggest that there are labeling mechanisms, both within and outside of Denmark, that have been specifically designed with SMEs in mind. In our work, we do not limit our exploration to labeling only, but take a comprehensive approach to investigate the different methods used by SMEs to demonstrate security.

\subsection{Further Security Practices in Companies}

A number of empirical studies have been conducted, investigating security practices within companies. The findings from these studies most relevant for our research are the role of trust in cybersecurity, internal communication, and security frameworks designed to support SMEs in particular.

\subsubsection{Role of Trust in Security}

Several works conclude the importance of trust in establishing security practices. As such, the study by Dalela et al on security practices in Danish companies~\cite{Dalelaetal} concluded that the prevalent trust within companies towards external partners, employees, and employers mirrors the broader importance of trust in Danish society as a whole. Tahaei and Vaniea \cite{Tahaei2019ASO} review practices related to use of third-party libraries, concluding that the trust developers have in these libraries is based either on these libraries being certified by standard organizations, or on their widespread use within the community.

\subsubsection{Internal Communication}

Jenkins et al. conducted a survey study with system administrators to investigate their self-reported behaviors towards patch management \cite{jenkinsetal}. The study found that factors such as resource availability, patching policies, and access to dedicated testing environments significantly influence their approach to testing. The study observed that patch management practices tend to be more ad-hoc in SMEs , in contrast to larger organizations, where the process is more systematic and structured. Thomas et al. conducted an interview study with 32 application security experts to explore their views on application security processes and their usability \cite{Thomasetal10.1145/3173574.3173836}. The study highlights a critical issue: while auditors are tasked with identifying security vulnerabilities in code, the responsibility for fixing these vulnerabilities falls on developers. This division of responsibilities can lead to communication challenges, which may in turn, compromise the security of the application. The solutions proposed emphasize the importance of organizational processes, support and training for developers, and automation of tools to reduce manual burdens. This runs on a parallel yet complimentary track to our work, as Thomas et al. investigate internal communication about security, while we focus on the external.

While these studies investigate how developers and security professionals interact with these tools, our study takes this a step further by examining how the outputs of these tools and related tasks can potentially be used to communicate the security status of an application or software to clients.

\subsubsection{Frameworks for SMEs}

To tackle the overwhelming nature of initiating work in cybersecurity for SMEs, some initiatives aim to simplify the process by e.g. breaking it into manageable and easily comprehensive steps. For example, the EU Digital SME Alliance recently introduced the SME Guide, based on ISO/IEC 27002, which prioritizes 16 essential controls to align with GDPR requirements \cite{digitalsme_guide_controls}. The ``SevenWeek'' method, evaluated by Gaitero et al. is designed to assist SMEs in creating or enhancing their quality and security management systems within a seven-week period. The results indicate that SevenWeek is both easy to implement and adaptable to the specific needs of small companies \cite{GAITERO2021110960}. Similarly, Ozkan and Spruit propose the ``Adaptable Security Maturity Assessment and Standardization'' (ASMAS) framework, which tailors security evaluations to the unique circumstances of individual SMEs. This framework not only aims to improve SMEs' security over time but also helps them benchmark their progress and communicate their security status effectively. While initial evaluations of the framework showed high scores for ``perceived usefulness'' and ``ease of use'', the framework scored lower on ``intention to use'', as some participants felt it was beyond their immediate needs \cite{OzkanandSpruit}. Tools like Sikkerhedstjekket (``security check'') provide tailored risk assessments and actionable recommendations but often fail to make this information easily accessible for SMEs \cite{virksomhedsguiden}, and Security Canvas \cite{securityscientist} incorporates contextual risk analysis inspired by NIST, and SMESec, a framework that emphasizes employee training and lightweight security measures. In summary, significant efforts have been made to support SMEs achieve compliance.

\section{METHODOLOGY} \label{sec:Methodology}

\begin{figure*}[t]
    \centering
    \includegraphics [width=1\textwidth] {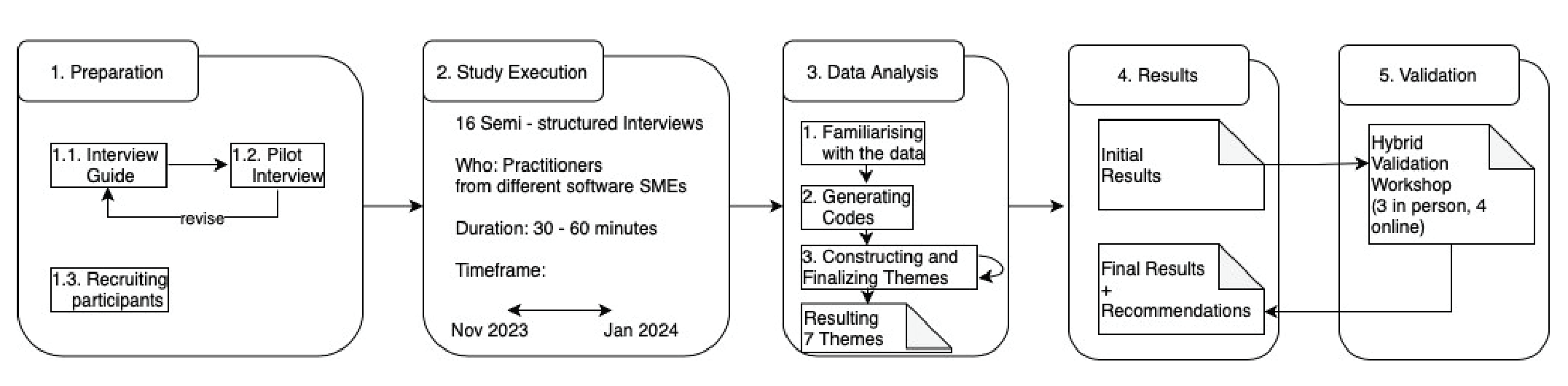}
    \caption{Study process overview}
    \Description{A diagram showing individual steps of the study process, namely (1) preparation, (2) execution, (3) data analysis, (4) results, (5) validation. }
    \label{fig:process}
\end{figure*}

To address the research questions presented in the introduction, we conducted ethnographic interviews to gain insights into the different methods and techniques used in industry to demonstrate security. The findings from the ethnographic interviews were validated in a workshop with a subset of the participants from the interview study. Figure \ref{fig:process} presents the methodology followed for this study. 

We describe each phase of our investigation in detail below.

\subsection{Qualitative Interviews - Preparation}

This section describes how the qualitative study was planned and conducted, and the methodology used for the interview data analysis.

\subsubsection{Participant profiling}

This study’s sampling covers SMEs as per the EU definition \cite{economic2024}, focusing specifically on SMEs engaged in software development. This includes small companies (10-49 employees) and medium-sized companies (50-249 employees). However, data from Statistics Denmark \cite{DanmarksStatistik2024} indicates that most companies in categories related to software development have fewer than 10 employees. Therefore, we have chosen to include SMEs of all sizes, encompassing micro companies (0-9 employees). We targeted companies operating under Danish industry classifications 'Computer programming', 'Other release of software' and 'Other IT service' — which primarily includes those whose core business is software development \cite{Nielsen}. However, companies where software development is a secondary activity but still significant were also considered within the scope of our project, such as 'Consultancy regarding information technology' and 'Other management consultancy activities'.

\subsubsection{Reach-out strategy and interviewee recruitment}
To recruit participants, a poster was designed with a brief project description, the said participant profile, and contact information. We used social media to attract potential participants to contact us for more information or participation. Everyone who expressed interest in participating was invited, as all fit the criteria, with the exception of two individuals who were excluded due to working in a large enterprise. We also actively recruited participants from our own professional networks, identifying those that would be suitable candidates.

In total, we conducted 16 interviews. The interviews took place from November 2023 until January 2024, and terminated when saturation was achieved, and no new findings were identified in the ongoing data analysis. The recruitment process was ongoing throughout the entire interview period. Table \ref{table:participant profiles} includes an overview of interview participants.

\begin{table*}[]
  \centering
  \begin{tabular}{| @{} l @{} | @{} l @{} | @{} l @{} | @{} l @{} |}
    \hline
    \hspace{5pt} \textbf{ID} \hspace{5pt} & \hspace{5pt} \textbf{Industrial Classification} \hspace{5pt} & \hspace{5pt} \textbf{Employees} \hspace{5pt} & \hspace{5pt} \textbf{Participant role} \hspace{5pt} \\ 
    \hline
    \hspace{5pt} P1 \hspace{5pt} & \hspace{5pt} \makecell[l]{Consultancy regarding information technology} \hspace{5pt} & \hspace{5pt} \makecell[l]{<10} \hspace{5pt} & \hspace{5pt} \makecell[l]{CEO} \hspace{5pt} \\
    \hline
    \hspace{5pt} P2 \hspace{5pt} & \hspace{5pt} \makecell[l]{Other release of software} \hspace{5pt} & \hspace{5pt} \makecell[l]{<10} \hspace{5pt} & \hspace{5pt} \makecell[l]{CPO} \hspace{5pt} \\
    \hline
    \hspace{5pt} P3 \hspace{5pt} & \hspace{5pt} \makecell[l]{Computer programming\\Consultancy regarding information technology} \hspace{5pt} & \hspace{5pt} \makecell[l]{50-249} \hspace{5pt} & \hspace{5pt} \makecell[l]{Compliance Manager} \hspace{5pt} \\
    \hline
    \hspace{5pt} P4 \hspace{5pt} & \hspace{5pt} Computer programming \hspace{5pt} & \hspace{5pt} 10-49 \hspace{5pt} & \hspace{5pt} \makecell[l]{Business Partner and \\VP of service operations}\hspace{5pt} \\
    \hline
     \hspace{5pt} P5 \hspace{5pt} & \hspace{5pt} Consultancy regarding information technology \hspace{5pt} & \hspace{5pt} <10 \hspace{5pt} & \hspace{5pt} Freelance Consultant \hspace{5pt} \\
    \hline
     \hspace{5pt} P6 \hspace{5pt} & \hspace{5pt} \makecell[l]{Business consultancy\\Other management consultancy activities} \hspace{5pt} & \hspace{5pt} 10-49 \hspace{5pt} & \hspace{5pt} DevSecOps Engineer \hspace{5pt} \\
    \hline    
     \hspace{5pt} P7 \hspace{5pt} & \hspace{5pt} Computer programming \hspace{5pt} & \hspace{5pt} 10-49 \hspace{5pt} & \hspace{5pt} Co-Founder \hspace{5pt} \\
    \hline
     \hspace{5pt} P8 \hspace{5pt} & \hspace{5pt} \makecell[l]{Computer programming\\Advertising agency} \hspace{5pt} & \hspace{5pt} 10-49 \hspace{5pt} & \hspace{5pt} CPO \hspace{5pt} \\
    \hline 
     \hspace{5pt} P9 \hspace{5pt} & \hspace{5pt} Computer programming \hspace{5pt} & \hspace{5pt} 10-49 \hspace{5pt} & \hspace{5pt} \makecell[l]{Product Manager\\IT and Security Compliance Manager}\hspace{5pt} \\
    \hline
     \hspace{5pt} P10 \hspace{5pt} & \hspace{5pt} Computer programming \hspace{5pt} & \hspace{5pt} <10 \hspace{5pt} & \hspace{5pt} CTO \hspace{5pt} \\
    \hline 
     \hspace{5pt} P11 \hspace{5pt} & \hspace{5pt} \makecell[l]{Computer programming\\Consultancy regarding information technology} \hspace{5pt} & \hspace{5pt} 10-49 \hspace{5pt} & \hspace{5pt} Developer \hspace{5pt} \\
    \hline 
     \hspace{5pt} P12 \hspace{5pt} & \hspace{5pt} Computer programming \hspace{5pt} & \hspace{5pt} <10 \hspace{5pt} & \hspace{5pt} CEO \hspace{5pt} \\
    \hline 
     \hspace{5pt} P13 \hspace{5pt} & \hspace{5pt} Other IT service \hspace{5pt} & \hspace{5pt} <10 \hspace{5pt} & \hspace{5pt} Security Expert \hspace{5pt} \\
    \hline 
     \hspace{5pt} P14 \hspace{5pt} & \hspace{5pt} Computer programming \hspace{5pt} & \hspace{5pt} <10 \hspace{5pt} & \hspace{5pt} CEO \hspace{5pt} \\
    \hline 
     \hspace{5pt} \makecell[l]{P15\\P16} \hspace{5pt} & \hspace{5pt} Consultancy regarding information technology \hspace{5pt} & \hspace{5pt} <10 \hspace{5pt} & \hspace{5pt} CTO and Senior/Lead Developer \hspace{5pt} \\
    \hline 
  \end{tabular}
  \caption{Table of interview participants}
  \label{table:participant profiles}
\end{table*}

\subsubsection{Interview structure planning}

Prior to conducting the interviews, a semi-structured interview guide was designed and developed. In the first half of the interview, our questions focused on understanding (1) the extent to which companies have requirements and constraints to demonstrate the security of their product or service, and (2) the ecosystem of requirements within the company, focusing on who imposes requirements on the company and how the company assesses the security of third-party. In the second half of the interview, the questions aim to cover the communication and demonstration practices related to security. The questions attempt to uncover tools, practices, and channels for conveying the security of the product or the service, as well as nudge the participant to provide specific examples and thereby reflect on the strengths and weaknesses of the handling and strategy. For more information and a complete interview guide we refer to Appendix \ref{appendix:a}.

Prior to conducting the interview sessions, we ran two pilot interviews with industry professionals, who represent the study's participant profiles, to validate our questions and identify flaws in the interview guide. All interviews were recorded with the participants’ permission and transcribed for analysis. To protect the confidentiality of the participants, their names and organizational details have been anonymized. The interviews took 52,6 minutes on average. The longest interview lasted 1 hour, 8 minutes, and 12 seconds, while the shortest interview was 28 minutes and 45 seconds.

\subsection{Interview data analysis }

The analysis of each interview was conducted using thematic analysis methodology \cite{BraunClarke2006}. For the classification of the collected data, an Affinity Diagram has been employed which aids in organizing extensive amount of qualitative data \cite{HoltzblattBeyer2017}. In an Affinity Diagram, all notes, in this case, quotes from the interviews, are transcribed onto individual post-it notes and then grouped into categories based on their similarities. An Affinity Diagram is thus built from the ground up, with no predefined categories that data must fit into, instead, the data is grouped into clusters that collectively form overarching categories \cite{HoltzblattBeyer2017}.

To classify the collected data, qualitative principles have been employed by categorizing the participant’s statements on the meaning conveyed \cite{BrinkmannTanggaard2010}. The classification process took place on the Miro platform~\footnote{https://miro.com}, where transcribed interview quotes were imported and allocated onto separate post-its. Subsequently, the quotes were grouped in similar thematic characteristics into smaller clusters and given the same post-it color. Once most quotes were grouped, related clusters with overarching thematic characteristics were merged and a headline for each category was devised. The quotes were iteratively reorganized to identify any gaps or potential misclassifications. 

A single researcher conducted the thematic analysis. In order to strengthen the reliability of the analysis, one interview was coded independently by two researchers, who discussed and presented the results to  the rest of the research group. Subsequently, the paper authors met regularly to review and edit the themes and analyze the data.

Based on the thematic analysis, we aimed to select key demonstration methods. In order to make sure that our selection of methods is representative of practices in companies, we decided to consider only the demonstration methods that were mentioned by at least 25\% of our participants (that is, at least 4 out of 16).

\subsection{Validation Workshop}

To validate our findings we extended a workshop invitation to all 16 participants, of which six accepted and participated in a validation workshop. The purpose was to present our key findings on security demonstration methods to the participants, and through interactive activities validate or challenge these findings and/or uncover new insights. 

Before the workshop, we conducted a pilot session, which served as both a rehearsal and a follow-up interview with an interview participant to ensure the smooth flow of the workshop and exercises. Their input has been integrated into the analysis alongside that of the other participants. The pilot workshop feedback include minor communication and cosmetic adjustments to the presentation, such as adding icons to the presentation slides, highlighting instructions in bold, and explaining the steps of the workshop in the introduction. The overall structure remained unchanged.

The format of the validation workshop was based on Liberating Structures, that are a set of activities fostering lively participation in groups of any size and kind and making it possible to include everyone and not just a few dominant voices~\cite{LipmanowiczMcCandless}. This method was deemed suitable for our purpose, as we wanted minimal interference and control over the activities. Given the time frame, we decided that two exercises would maintain a progressive flow in the workshop while allowing time for organic group discussions. The workshop consisted of two overarching exercises aimed at fostering discussion about security demonstration methods, separated by a 15-minute break. We provide more detail on each individual exercise and its structure and results in \ref{sec:workshop results}.

\subsection{Ethical considerations}

Although our institution does not have mandatory Institution Review Board for studies, we were required to submit the privacy impact assessment specifying our plans for collecting and managing data, which was reviewed by the data protection officer of our institution. Furthermore, we adhered to the common practices of addressing ethical considerations~\cite{BrinkmannKvale2017}. Throughout our study we were conscious about applying the ethical principles of confidentiality to safeguard the privacy of participants’ data. As such, we followed established guidelines when conducting the study, in line with the General Data Protection Regulation (GDPR), as well as assuring participants that their personal data would remain confidential and that the results will only be reported in anonymised format. Accordingly, prior to participation, we explicitly informed participants  about the purpose of the study and their voluntary participation was ensured through informed consent. We did not provide any remuneration to our interviewed participants.

\section{Interview Results} \label{sec:Interview Results}
In this section, we discuss the results of the qualitative interviews. We begin by describing where requirements for security demonstration come from, followed by the five key methods for security demonstration that emerged from the analysis of the interviews. From our analysis, the identified key methods were mentioned by 7 out of 16 of our participants. These findings are detailed further in \ref{sec:Interview Results}.

Mentioned methods excluded from our results were omitted for one of two reasons: (1) they were not considered methods of security demonstration because they lack the capacity to actively provide assurance of the security in place, such as contractual agreements (e.g. data processing agreement and general best practices), and therefore not in line with the focus of the study or (2)  they were unique to individual companies, e.g., using illustrative drawings or referring to company specific Wikipedia pages, and therefore lacked generalizability. 

Results are described in detail below.

\subsection{[RQ1] Security requirements}

Our interviews revealed that the security requirements that companies had to comply with were mainly imposed by their \emph{clients} or demanded by \emph{legislation}. The requirements themselves also varied, depending on their origin, with some being strictly mandatory and requiring precise adherence and others more flexible and open to negotiation. We elaborate on requirements as originating from these different sources (clients vs legislation) below.

\subsubsection{Clients}

Our participants suggested a division between large clients imposing more stringent requirements and small clients with flexible or minimal requirements. 

\paragraph{Large clients} Companies interviewed note a mix of clients comprising mainly larger clients, followed by SMEs and a few public institutions. Larger clients maintain focused cybersecurity efforts, with designated teams and specific security requirements, including certifications and supplier due diligence. Our participants state that larger clients typically involve their IT department when on-boarding new suppliers, considering it a standard process. The security requirements originating from large clients include specific certifications (such as IEC 62443, SOC2, ISO 27001), penetration test reports, audit reports (such as ISAE 3000 or 3402), security questionnaires and socio-technical requirements incorporated into contracts (such as using MFA and encryption, not hosting on servers in certain geographic regions, and clients conducting their own penetration tests). Our interviewees furthermore noted that these requirements can vary widely, with some clients only seeking minimal documentation (such as documentation of compliance with a standard through e.g., self-assessment, confirming that an organization adheres to a specific set of guidelines but without necessarily obtaining the formal certification) and others requiring a more detailed evaluation (such as asking for updated audit or test reports annually) before engaging in business.

\begin{quote}
    \textit{``I would say about 1/3 of our clients currently ask for updated reports on an annual basis.''} [P4] 
\end{quote} 

\begin{quote}
  \textit{``(...) the framework of ISO 27001 would be something that anytime we talk to financial institutions they would ask us for that framework. Maybe not necessarily the actual stamp of approval from an accountant, but they would ask us to document our processes according to that framework.''} [P12]
\end{quote}

\subsubsection{Small clients}
Interviewed companies noted smaller clients often have minimal or no specific security requirements. This lack of demands is attributed partly to these clients' limited understanding of cybersecurity complexities and their lack of dedicated expertise in the field. Additionally, their financial constraints and organizational scale may limit their ability to present requirements and ultimatums. 

One participant mentions the knowledge gap among SME clients regarding their awareness of applicable legislation, which forces them as suppliers to be proactive in establishing necessary security agreements with these smaller clients. Another highlights that smaller companies lack technical staff. One participant [P7], acting as a third-party supplier in a supply chain, provides an example where a small client requested security demonstration through provision of relevant certification due to demands from their own clients. P7 explained that these requests were declined, and in response, the participant suggested charging their SME client for certification or documentation provision, as it's not economically feasible to engage in such activities solely for individual clients.

\begin{quote}
    \textit{``(...) But they basically don’t pay us enough that we have prioritized to make an effort just for them. So we have told that if they need like certifications or revision statements they would have to pay us to make it, because there isn’t a sustainable business for us in doing that for them alone.''} [P7] 
\end{quote}

When smaller clients do specify requirements based on enforced legislation, like GDPR, they may use templates provided by governmental agencies like, in this case, The Danish Data Protection Agency to help ensure compliance to GDPR. Similarly, another participant noted that smaller clients often rely on standard services, such as those provided by Microsoft, assuming that if the supplying company uses these services, the proper level of security is ensured.

\begin{quote}
    \textit{``The small organizations just basically rely on the Microsoft ecosystem to be secure, you know, if we use those products and we follow their guidance and so forth and that's fine. That's, I think, the attitude I meet, right.''} [P8]
\end{quote}

\subsubsection{Legislation}
All interviewed companies are situated in Denmark, Europe, thus they are to some extent subject to European legislation. Whether a company is compliant with the legislation will be revealed through audits. As an example, one participant explains that their clients impose audit requirements and, in the audits, they have a designated section that addresses how personal data or financial data is handled. Another participant, that services airports globally, highlights their reliance on clients to ensure compliance with country-specific legislation, and cites the infeasability of keeping track of all legal requirements across countries and jurisdictions. In other words, companies adhere to legislation that is supervised by their client(s). 

Legislation appears clear in certain instances, yet vague and open to interpretation in others. A participant asserts that developers face challenges due to this ambiguity, because written laws offer some clarity, but certain aspects remain subject to interpretation, so uncertainty persists until legal boundaries are established through court cases.

\begin{quote}
    \textit{``You could look at the legislation but then there’s a legal interpretation, and you don’t really know whether you break that before the case has been tried at the courts, right? Because, I mean, the legislation is one thing and... But the real boundaries of the legislation is not set down before things have been tried in courts. So, I think that for developers it’s kind of a gray area where...Are you compliant? Are you not compliant? Because it’s a legal interpretation that you have to call a very expensive lawyer to come with their interpretation of the law and even very expensive lawyers do not agree, right?''} [P11]
\end{quote}

\subsubsection{RQ1 Summary}

To summarize where security requirements originate from, our participants reported the following: 

\begin{itemize}
    \item Large clients have designated cybersecurity focus and specific security requirements, but the extent of the requirements they impose on their suppliers can vary. The requirements can include obtaining specific certifications, filling out questionnaires, being subject to audit reports and penetration tests, but may vary in levels of documentation required based on the client. 
    \item Small clients present minimal requirements, and if they do, their organizational scale and significance are insufficient to have an impact on what SMEs decide to do to satisfy those requirements. Small clients may rely on resources, such as templates, provided by governmental agencies, especially when specifying requirements based on enforced legislation.
    \item Companies are subject to legislation, as such they have to comply to requirements set by lawmakers. Additionally, specific legislation requirements can be specified by clients. 
\end{itemize}

\subsection{[RQ2] Methods of Security Demonstration}
\begin{figure*}
    \centering
    \includegraphics[width=1\textwidth]{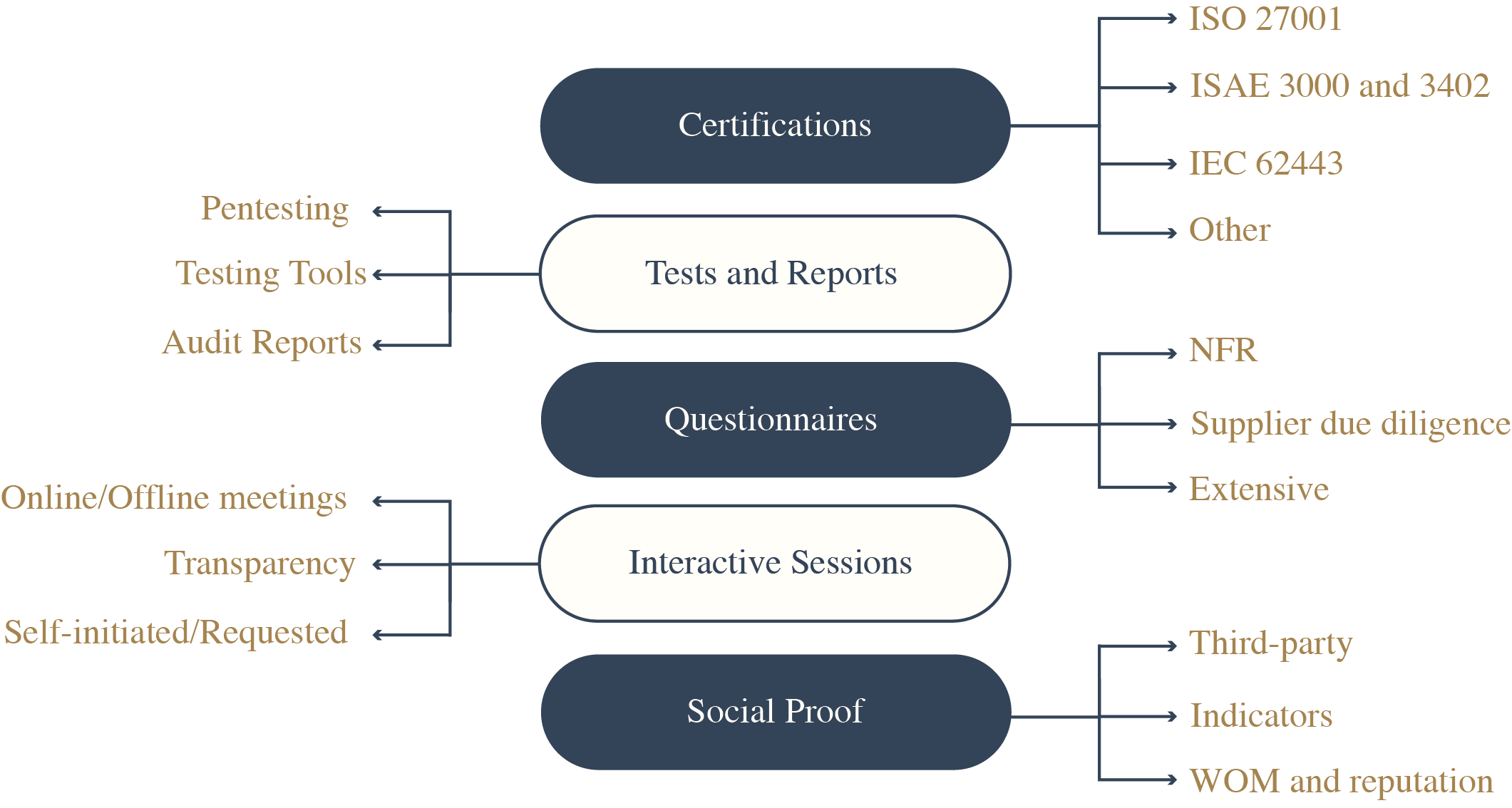}
    \caption{Five key methods for security demonstration and what they each encompass}
    \Description{An overview of the five key methods for security demonstration}
    \label{fig:5keymethod}
\end{figure*}

Our results indicate five key methods for security demonstrations shown in Figure \ref{fig:5keymethod}.

We describe each in detail in the following subsections.

\subsubsection{Certifications}

Certifications refer to standardized national or international certifications obtained and maintained, through audits, by companies. Our participants mention that the certification requirements imposed on them vary in rigidity. In some cases, they are non-negotiable prerequisites for collaboration, such as providing ISAE 3402\footnote{An international standard, specific to service organization engagements, governing controls over financial reporting to provide assurance regarding internal controls, hereunder ensuring the confidentiality, integrity, and availability of financial data.} or 3000\footnote{An international standard that provides assurance over non-financial information, including but not limited to GDPR.} on an annual basis or certification documentation according to SOC2\footnote{An information security and audit framework specifically designed for service providers storing customer data in the cloud, and relates to assurance on IT controls.}, IEC 62443\footnote{An international standard providing guidelines for cybersecurity in industrial automation and control systems throughout their lifecycle.}.

\begin{quote}
   \textit{``We had one customer who actually liked the platform a lot, but he said their DPO said if we did not have a SOC2 or similar then he was not allowed to buy our services.''} [P2]
\end{quote}

In other instances, participants note that clients may request certifications but offer flexibility in negotiations, allowing companies to discuss terms or substitute with alternative security demonstration methods. As an example, one participant highlights that some larger clients would like them to have an ISAE 3402 declaration, but it is something that they are trying to avoid due to the associated financial burden that is not in line with the company’s resources.

Several participants echoed that the price of a certification is one of the towering barriers to embrace, and in some cases, obtaining a certification may not be realistic at all for an SME with limited resources. Similar concerns evolve around the level of bureaucracy that comes along with being certified and having to undergo regular audits, which results in changing and/or implementing internal processes without any immediately evident value. In some instances, standards are used to demonstrate compliance even without seeking a formal certification, as one participant elaborates:

\begin{quote}
  \textit{``We have just finished our ISO 27001 \footnote{An international standard for establishing, implementing, maintaining, and continually improving an information management system (ISMS) within an organization.} project where we managed to get a compliant check, we're not certified. That's simply too expensive and the problem is when you're first are certified don’t ever be uncertified because then people will say, oh you've gone down. Whereas when you say you're compliant and you can present the documents, you can present the findings and the ongoing reports then people are happy.''} [P2]
\end{quote} 

The interviewed participants debated the reliability of certifications, arguing that certifications do not provide absolute security assurance. They are indicators of security, however, there is still a certain degree of trust that applies.

\begin{quote}
    \textit{``(...) they are certified for SOC2, they do say that they apply best practices in their secure coding, but then you have to trust that they actually do it when they develop. That's not something I think we could even check for that.''}[P13] 
\end{quote}

Another vital feature of a certification is its relevance for the ecosystem a company operates in. As an example, P6, the software and consultancy company servicing airports globally, feels challenged by the legal implications of every country, and even though most requirements and certifications have similar characteristics across the world, there are still legal differences, preventing certifications from being absolute guarantees.

\begin{quote}
    \textit{“Again, it's those specific countries’ laws which can differ incredibly much. That is just impossible to keep track of. So I would say yes, standards help, but that's not the silver bullet or the whole thing.”}[P6] 
\end{quote}

\subsubsection{Tests and Reports}
Reports of different kinds of internal security tests and audits are furthermore used for security demonstration. As such, several participants mention using reports of penetration tests employed to assess and reveal vulnerabilities in their security, as security documentation provided upon request. These tests can be conducted automatically or manually by the company's employees or external parties. 
Participants describe conducting internal penetration tests as a common practice, automated to simulate attacks on all platforms, systems, and ports, ensuring they are updated or notified on areas that need updating. One participant mentions that external parties conduct penetration tests annually on their software, reporting identified vulnerabilities and their risk levels, alongside a report which can be used as documentation upon client request. Participants also reported on receiving requests from clients wanting to conduct their own penetration tests. 
Some utilize software to perform automated penetration tests, generate test reports, and indicate security levels on colour scales, including mentions such as SonarCloud\footnote{https://www.sonarsource.com/products/sonarcloud/ 02/09/2024} and SonarQube\footnote{https://www.sonarsource.com/products/sonarqube/ last visited 02/09/2024}, with the advantageous ability to immediately generate reports to be used for security demonstration. 

\begin{quote}
   \textit{“They could do that but then like we would not perform a penetration test at that time, we would give them the newest report. But then they can again perform penetration tests of their own anytime they want. They just have to give us a heads up, obviously, if they’re gonna attack us.”} [P6] 
\end{quote}

However, while participants mention that tests can be a promising tool for security demonstration, they also emphasize that they mainly provide a brief security snapshot, as new vulnerabilities may emerge shortly after a test has been conducted, making the security obsolete, or not paint a comprehensive security picture as they are modified to test specific parts. In addition, they are costly to conduct.

\begin{quote}
  \textit{“But pentest is also kind of a one shot, you release software and you pentest it. But I mean an hour after you can have released it, there can be a new vulnerability and then the pentest is obsolete. So the pentests, they are good requirement but they're not that efficient and they're very, very expensive.”}[P5]
\end{quote}

Similar to penetration test reports, which summarize a system’s vulnerabilities, including patching and improvement recommendations ~\cite{browserstack2024}, audit reports are also common security demonstration methods. Audits reports are assessments of an organization’s information systems’s security against externally established standards or federal legislation ~\cite{auditboard2024}. While some audit reports can be made accessible via a company's website or system, others are restricted from marketing use according to agreements with the auditing firm.

\begin{quote}
    \textit{``So once we get audited, we cannot use it in our marketing because we paid the third party to audit us.''} [P16]
\end{quote}

\begin{quote}
     \textit{``Inside our system they can like always download our two audit reports (...)''} [P10]
\end{quote}

In sales discussions, referencing a recent audit report can instill confidence. Many clients perceive security upon seeing established auditing firms, often eliminating the need for further reassurances.

\begin{quote}
    \textit{“So they're basically satisfied once you have the audit report from, we use PwC but it could have been KPMG or anyone else. They basically just reach their conclusion. I think it's sort of, it seems that way anyway. So obviously in the sales material, the sales pitch, they will all ask questions and we'll go over it and like adapt the settings on how long to keep data after refusal, after approval but in the end, they'll just be like ‘Yeah, OK. If PwC says it's OK, it's OK.’”} [P10]
\end{quote}

\subsubsection{Questionnaires}
Questionnaires consist of standardized yet highly detailed questions designed to assess a company’s security procedures. Large clients, in particular, require suppliers to complete these questionnaires. The questionnaires can consist of anything between 75 to 300 questions, but the specific inquiries may vary from one client to another. 

The consensus among the interviewed participants is that these questionnaires address security both at the surface level and delve into specific details. Questionnaires can be a supplement to certifications or audit reports for security demonstrations, or they can be a substitute for it. As an example, if a third-party vendor is unable to provide an ISAE 3000 report a questionnaire substitutes to assess their security management by inquiring about IT policies, access policies, background checks, security documentation and confidentiality. Questionnaires can also be utilized by the respondent company to identify internal areas that have adequate security and areas demanding further attention. One participant reflects on an experience where providing detailed answers was viewed favorably by the client, enhancing their posture as a company among their competitors as well as getting assurance on their security practices.

While questionnaires have the potential to serve as valuable instruments for both internal and external security assessments and collaboration, they can become resource-intensive, by necessitating companies to allocate significant resources to respond to extensive questionnaires, especially when facing similar yet not identical inquiries from various parties. Interviewed participants view this process as repetitive and demanding, ultimately preferring certification as a more efficient solution. Furthermore, upon closer examination of questionnaires, companies may find it challenging to provide satisfactory responses that align with the intended purpose of specific questions. This ultimately portrays the questionnaire as generic, primarily geared toward compliance rather than genuine security.

\begin{quote}
    \textit{"We have also experienced that some customers have sent a lot of IT security related requirements in a questionnaire and then we take this very seriously and we go down and answer all of the questions and if we meet questions that we don’t understand, we will try to get some information from the customer saying, 'OK, what do you mean by this?'. And we have experienced actually multiple times when we sort of push back a little bit to the customer, they will say 'ohh shit we don’t even know what we’re asking'. So it sometimes we get a feeling that all of these questions that the customers sometimes ask us, it’s just like a... Because they have to. It’s like a requirement that they have to from some standard, but that they don’t actually know exactly what they’re asking us, which is a little bit strange.”}[P9]
\end{quote}

\subsection{Interactive Sessions}
Interactive sessions are physical or virtual meetings, during which the participants report addressing security visually and personally. These sessions involve direct dialogue between the company and the client, serving as either the audit method itself or to address follow-up questions. 

Typically, during these interactive sessions, the supplier company will open its systems, repositories, and code files and respond to the client’s specific questions by referring to the current state. Participants report that in such sessions, a positive perception of transparency often emerges. Self-initiated interactive sessions are also viewed as advantageous by several participants, facilitating visual discussions and adaptable communication strategies. However, there are also mentions of challenges in maintaining professionalism while being fully transparent about potential vulnerabilities or errors.

\begin{quote}
\textit{"What can be challenging is that we're exposing some of our internal mechanisms like some our internal way of working in front of the customer and of course we try to do it in the most professional way, but we also have to accept that we are dealing with our technical team, which are human beings and these human beings sometimes also can write code in different ways (...)”} [P9] 
\end{quote}

\subsubsection{Social Proof}
Participants describe social proof as a method they use to assess security when implementing third-party software. They mention looking at indicators such as ratings, number of users, last update, track record, community usage, and reputation.

Some participants argue that open-source software's security is reinforced by its wide examination, while others caution against relying solely on this aspect. Some emphasize community feedback for identifying vulnerabilities, others advocate for additional measures like developer training and rigorous testing of software libraries. Participants also express social proof by selecting partners or providers with widespread usage and recognition within their sector. Reputation plays a significant role in their decision-making process. 

\begin{quote}
\textit{“Well, typically our clients discuss with us what our solution is, how it's built, and since it contains open-source components, they know that.. They know the security level of these components, and after discussions with this, most clients are happy with how the product is structured.”}[P1]
\end{quote}

Testimonials and bench-marking represent another influential form of social proof. When a company showcases compliance with the stringent security standards demanded by top-tier clients, it significantly reinforces trust in their implemented measures.

\begin{quote}
   \textit{“So they all ask for the same and that's why we have from the beginning invested a lot in being able to demonstrate compliance and security. First when you can fulfill the requirements for those that have the highest security requirements, then everybody else is just happy that you do.”} [P3] 
\end{quote}

\subsubsection{RQ2 Summary}

To summarize the key security demonstration methods, our participants reported the following:
\begin{itemize}
    \item Certifications: while recognized as effective, come with significant acquisition and maintenance costs.
\end{itemize}
\begin{itemize}
    \item Tests and Reports: whether conducted internally or externally, manually or automatically, serve as a valid demonstration tool.
\end{itemize}
\begin{itemize}
    \item Questionnaires: often described as extensive, are provided by clients and answered by the supplying company.
\end{itemize}
\begin{itemize}
    \item Interactive Sessions: involve direct interaction between the supplying company, client, or auditing firm, addressing security inquiries or concerns.
\end{itemize}
\begin{itemize}
    \item Social Proof: based on factors such as community usage and personal evaluations, also play a significant role in assessment. 
\end{itemize}
\section{Workshop Results} \label{sec:workshop results}
In this section, we discuss the detailed structure and results of the validation workshop.

Our participants are situated in diverse geographic regions of Denmark, therefore, we offered hybrid workshop participation. We facilitated the physical workshop on March 25th, 2024, and online participants joined via Microsoft Teams that was connected to the meeting room’s designated TV screen, enabling all participants to engage in dialogues during the workshop exercises. The workshop was scheduled for 2,5 hours. Three participated physically (P1, P7, P12~\footnote{Participants are referenced by the same ID in both the workshop and interview results sections of this paper.}) and three participated online (P4, P5, P9), and we conducted the pilot session with P12. 

We used paper, post-its and writing utensils to record the ideas of physical participants, and we intended to use a Miro board for online participants. However, one participant joined via audio, and another had platform access issues in Miro, so they documented notes in Teams chat and completed the activities verbally. Insights were photographically recorded as they emerged from exercises. The workshop was audio recorded with the participants’ permission through a consent form and then transcribed for analysis. To protect the confidentiality of the participants, their names and organizational details have been anonymized.

As mentioned in the Methodology section (Section~\ref{sec:Methodology}), the workshop was based on Liberating Structures, and the first exercise was based on the 1-2-4-All exercise~\cite{liberatingstructures_1-1-2-4-all}, where we aimed to present participants with the five key methods for security demonstration derived from our interview results. By asking them to reflect on the strengths and weaknesses of these methods, we sought to capture their immediate thoughts and validate the results from our interview analysis or discover new insights. The second exercise utilized the Min Specs structure~\cite{liberatingstructures_14_min_specs}, involving participants in devising recommendations for demonstrating software security. The Min Specs exercise focuses on identifying the minimum number of essential rules by eliminating non-essential elements. In other words, the outcome of this exercise is a concise list of must-dos and don’ts for selected security demonstration methods.

In the following subsections we will describe the detailed structure of each activity followed by the results and discussions.

\subsection{Activity 1}

The first main activity was inspired on the `1-2-4-All' structure - adapted to accommodate our number of participants - where we asked questions in response to the presentation of key findings, allowing participants to first, reflect in silence, then share their ideas with the other group members (3 people), and finally discuss and present in plenum\cite{liberatingstructures_1-1-2-4-all}. The first exercise had three iterations, each with different questions but with the same sub-questions and structure. The main activity questions for each iteration were as follows:

\begin{enumerate}
    \item From the presented themes, select 1 or 2 security demonstration methods that you have used in the past and liked 
    \item From the presented themes, select 1 or 2 security demonstration methods that you have used in the past and disliked 
    \item Can you imagine any other alternative methods to demonstrate security? 
\end{enumerate}

We then asked participants to silently reflect on at least 3 advantages and 3 disadvantages for the chosen demonstration method, share their ideas with the other group members, followed by a group discussion and list expansion, if needed. 

\subsubsection{Activity 1.1}For the first iteration, answering the question "From the presented themes, select 1 or 2 security demonstration methods that you have used in the past and liked" the participants discussed the following methods and their advantages and disadvantages: \\

\textbf{Certifications:}\\ 
As confirmation of our findings from the interviews, following and implementing the compliance framework, without necessarily acquiring the official stamp, can be considered valuable as it gives a foundation for internal security practices. In particular, the participants of our workshop said that they follow and implement a standard as it gives them a foundation for security and supports the dialogue with clients.
As such, it can reduce expenses for external auditing, arguing that presenting the documentation package to clients can suffice as security demonstration. Certifications also cover a wide spectrum of issues, serving as a structured approach to security and are thereby established as recognized method for security demonstration. However, they are resource-intensive and disproportionately costly for SMEs to obtain and maintain. During plenary discussions, and as new insights, participants debate the extent to which certifications, at present, comprehensively cover all aspects of security. However, it was acknowledged that even if they do not, certifications still hold value as they serve as descriptive standards outlining necessary requirements. \\

\textbf{Tests and Reports:} \\
As confirmation of our findings from the interviews, there are some obstacles to undergoing audits and penetration tests, that include cost-related concerns and time constraints. One participant argues that penetration test reports only paint a brief security picture. As new insights, participants discussed that audit and penetration tests reports offer standardized security approaches. While audit reports follow a more structured methodology, penetration testing can be ad-hoc in terms of what is being tested and how it is being tested. Therefore, it is crucial that the receiver of penetration test reports understands testing objectives and approaches. Testing, receiving feedback, and integrating it into teams is complex and time-consuming. Another participant says that publicly sharing their testing procedures on their website is an affordable and effective security demonstration method.\\

\textbf{Interactive sessions:} \\
While only briefly mentioned, the advantage of interactive sessions lies in their ability to provide a deeper understanding of client concerns, which is a confirmation of the interview findings. As new insights, participants discussed the unstructured and undocumented nature of interactive sessions as a security demonstration method. Additionally, it was mentioned that in certain setups, an informative video can substitute an interactive session.

\subsubsection{Activity 1.2}For the second iteration, answering the question "From the presented themes, select 1 or 2 security demonstration methods that you have used in the past and disliked" the participants discussed the following methods and their advantages and disadvantages: \\

\textbf{Questionnaires:} \\
While workshop participants disliked security questionnaires, they acknowledged their advantages. Questionnaires can substitute for unavailable certifications, provide a competitive edge, and serve as internal audits to reveal system gaps. However, they criticized the inconsistency, irrelevance, and complexity of questions, often crafted by external parties, making them lengthy and time-consuming. All of these confirmed our interview findings, but one participants noted that in their experience, action points often lose focus after completing the questionnaire, which is a new insights. \\

\textbf{Social proof:}\\ 
Briefly mentioned, but confirming our interview findings, social proof mainly involves seeking or providing referrals and recommendations, which can foster trust, but navigating its subjective nature can present challenges in the decision-making process, and one participant refers to it as an unstructured and undocumented method, so it should primarily be viewed as a contributing factor. As a new insight, participants argue that in certain cases, companies can freely exchange referrals and information, while in others, confidentiality constraints may limit such interactions.

\subsubsection{Activity 1.3} In response to the third question, participants got the creative freedom to propose alternative ways to demonstrate security, drawing on their own discussions and/or inspired by the five demonstration methods presented during the workshop.
 \\

\textbf{Framework toward SMEs:} A tailored template/framework for SMEs in software development or supply chains that includes minimum requirements and guidance from the start.\\

\textbf{Automated penetration tests:} An easy-to-use solution valuable for security on a practical level. It will be a cheaper alternative to regular penetration tests, which are expensive and only cover certain areas. The disadvantage identified is that such a tool will not cover zero-day attacks.\\

\textbf{Security culture:} The importance and combination of security tools (such as SCA, SASD, DASD and cloudified tools) and establishment of a security culture (e.g. through programs) for the software developers. \\

\begin{figure*}
    \centering
    \includegraphics[width=1\textwidth]{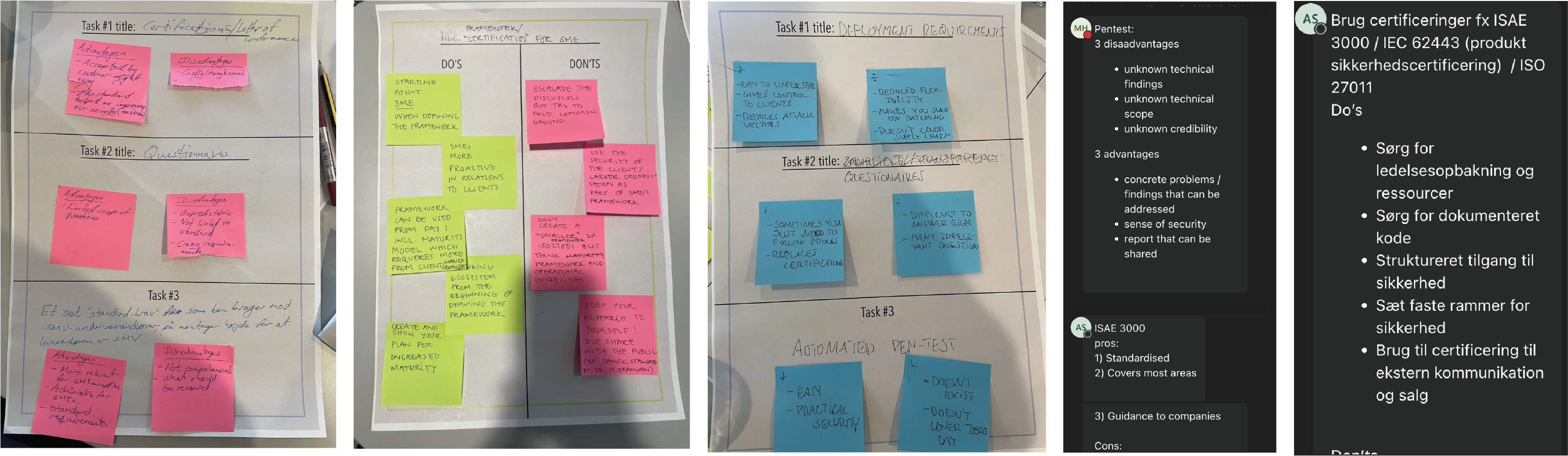}
    \caption{Photograph of workshop notes}
    \Description{Workshop notes showing post-it notes (for onsite participants) and screenshots from the chat (for remote participants) outlining the discussion outcomes}
    \label{fig:workshopnotes}
\end{figure*}

Figure \ref{fig:workshopnotes} includes photographs and screenshots of notes of the ideas that emerged during the workshop across all activities.

\subsection{Activity 2}
In this exercise, participants were asked to think of the minimum number of simple rules by eliminating the clutter of non-essentials, with the outcome containing a list of must dos and don'ts for selected security demonstration methods. 

\begin{enumerate}
    \item Pick a method that you will be most interested in using, either start using or continue using (1 minute) 
    \item If someone else were to use this method, what advice or recommendation would you give? Make a list of at least 5 dos and 5 don'ts (2 minutes)
    \item Discuss your notes in pairs/groups, expand the list if needed (4 minutes)
    \item Condense the list to its essential elements by asking yourselves "if we broke this rule, could we still achieve our purpose?". Remove the post-its that are unnecessary. (5 minutes)
    \item Present to everyone main discussion points. \\
\end{enumerate}

The physical participants requested to perform this activity as a group exercise, as they wanted to further develop their idea, \textit{Framework toward SMEs}, from activity 1.3. We discussed with them potential concerns about bias by explicitly asking whether group discussions might influence their responses. They reassured us that the discussions enhanced, rather than compromised, their ideas. Given that the workshop structure resembled a semi-structured interview, we prioritized flexibility over rigid adherence to the initial plan, allowing room for manoeuver, and not following the initial Liberating Structure by the book. The other participants independently focused on \textit{certifications} and \textit{interactive sessions}. 
The following lists summarize the participants' post-it notes and the discussions derived from the workshop audio transcript.\\

\textbf{Framework toward SMEs: DOS}
\begin{itemize}
    \item Framework should be developed with SMEs in mind.
\end{itemize}
\begin{itemize}
    \item SMEs should present their documentation packages willingly rather than waiting for requests. The proactive approach involves confidently addressing security measures before they are questioned, as inquiries are inevitable.
\end{itemize}
\begin{itemize}
    \item The framework should be seamlessly integrated into SME processes from day one, with built-in reminders for development stages and a maturity model to accommodate varying levels. It should also establish requirements for client engaging with SMEs.
\end{itemize}
\begin{itemize}
    \item To develop a solution that supports SMEs within larger ecosystems, ensuring their inclusion and addressing the unique challenges they face. This approach would also mean imposing additional requirements on larger corporations as recipients within the ecosystem.
\end{itemize}
\begin{itemize}
    \item The framework should allow SMEs to gradually improve their security measures as they grow, ultimately achieving certification.
\end{itemize}
\begin{itemize}
    \item It should encourage SMEs to use the customer’s security infrastructure when the software runs on the customer’s premises. For example, if the client already has ISO 27001 certification for their server room, it may be beneficial for the SME to host their software there rather than in their own server room.\footnote{This was placed as a \textit{don’t}, but during discussion it was discussed as a \textit{do}}
\end{itemize}

\textbf{Framework toward SMEs: DON'TS}
\begin{itemize}
    \item Avoid escalation in SME-to-large company dynamics in regards to security, by maintaining positive and open interaction and seeking common ground.
\end{itemize}
\begin{itemize}
    \item Don't downsize an already existing framework, but tailor it to the SME segment from the start. 
\end{itemize}
\begin{itemize}
    \item Don't keep your research to yourself, but share your insights with the public as collaborative efforts are crucial for achieving wide-spread adoption. \\
\end{itemize}

\textbf{Certification: DOS}
\begin{itemize}
    \item Get management support and allocate resources. It is essential to prioritize security early on to ensure more efficient coding practices and prevent issues in production.
\end{itemize}
\begin{itemize}
    \item Code documentation and documentation of code libraries. Even if they are open-source they might have unlucky side effects or uninteded consequences. 
\end{itemize}
\begin{itemize}
    \item Establish fixed frameworks and utilize them to get a structured approach to security.
\end{itemize}
\begin{itemize}
    \item Utilize certification as an effective communication and sales tool, ensuring its relevance by aligning it closely with the context of the organization’s offerings, rather than relying solely on broad certifications like ISO 27001.
\end{itemize}

\textbf{Certifications: DON'TS}
\begin{itemize}
    \item Don't just pick an easy certification. Consider certification, or at least adopt a structured approach to security, and make use of ISO 27000 guidelines for documenting information security management systems (ISMS) and security procedures, even if not pursuing full certification, to benefit from structured risk assessment and asset management processes.\\
\end{itemize}

\newpage
\textbf{Interactive Sessions: DOS}
\begin{itemize}
    \item Ensure that the meeting schedule allows ample time for discussion and that individuals with relevant expertise are in attendance. Specifically, those with a deep understanding of the rationale behind the questions being posed and the technical implications should be present.
\end{itemize}
\begin{itemize}
    \item Give feedback after session and make sure all parties are aligned on agreements and goals set at the meeting.
\end{itemize}
\begin{itemize}
    \item Receive specific customer cares/concerns before meeting to be prepared to address specific concerns thoroughly.
\end{itemize}

\textbf{Interactive Sessions: DON'TS}
\begin{itemize}
    \item Don't rush the session. Attentiveness is crucial during interactive sessions, as there may be underlying reasons motivating the client’s request for engagement.
\end{itemize}
\begin{itemize}
    \item Express lack of sympathy on their concerns, but demonstrate genuine concern for the customer’s concerns to foster trust and a personal connection. By doing so, they will perceive you not just as a representative of the company but as someone they can rely on.
\end{itemize}
\begin{itemize}
    \item Don't run the session until the customer has signed and agreed to pay extra for the interactive session. These sessions should exclusively involve existing customers who are willing to invest extra for this personalized service.
\end{itemize}

\subsection{Workshop Summary}
The validation workshop aimed to assess the extent to which our findings from the semi-structured interviews were supported or challenged. During the two main exercises, we presented our key findings and asked participants to consider them from various angles, either reinforcing or questioning the results. Overall, our findings were largely confirmed, but the workshop revealed additional nuances, particularly in methods like \textit{Interactive Sessions} and \textit{Social Proof}, which were described as more informal practices by the interview participants.

To summarize the key workshop findings, the results point to the following:
\begin{itemize}
    \item Certifications: offer a structured approach to security, and are also recognized as effective security demonstration methods. Participants highlight the efficiency of implementing compliance frameworks even without an official stamp, but emphasize the importance of ensuring that the certification aligns closely with the context of the organization's offerings. However, certifications are a costly undertaking.
\end{itemize}
\begin{itemize}
    \item Tests and reports: conducted either structured or ad-hoc, posit that the recipient of a test or audit report has adequate knowledge about the subject, and they can also incur significant costs. But they are recognized as effective, as e.g. audit reports typically follow a more structured methodology. So if such reports are made publicly accessible, they can be an affordable method to demonstrate security.
\end{itemize}
\begin{itemize}
    \item Questionnaires: often overwhelming in scope with irrelevant details, offer an alternative to certifications and provide the supplying company with the opportunity to utilize it to their advantage, as long as action points for areas of improvement do not lose focus after completion. 
\end{itemize}
\begin{itemize}
    \item Interactive sessions: are advantageous as they provide deeper understanding of client concerns, however, their unstructured and undocumented nature can be a disadvantage.
\end{itemize}
\begin{itemize}
    \item Social proof: acknowledged for its subjective nature, is recommended as a valuable contributing factor.
\end{itemize}

\section{Discussion} \label{sec:discussion}

Our results show that companies use a variety of methods for demonstrating the security of their products, which all have their advantages and disadvantages. As such, several of the methods can be considered \emph{passive} ones, not requiring an explicit interaction with the client to demonstrate security (namely, \textit{certifications, tests and reports} and \textit{social proof}). Such methods, however, can lack in flexibility, i.e. being able to adjust the demonstration towards specific security requirements of a client or the client's level of technological expertise (e.g. penetration tests posits that the receiver has the technical expertise to interpret them). On the other hand, \emph{active} methods that actively involve the client in the demonstration (namely, \textit{questionnaires} and \textit{interactive sessions}) can provide such a flexibility, but they can also be time-consuming due to the additional interactions required from both the client and the company. Choosing an appropriate demonstration method thus requires a consideration of a variety of factors: 

\paragraph{Costs:} Costs, monetary or related to other resources, were often mentioned as an obstacle for different security demonstration methods. As such, while \textit{certification} has often been mentioned as a desirable, or in some cases, non-negotiable requirement, its cost has been mentioned as the main barrier towards getting certified. At the same time, once obtained certifications can be used for security demonstrations with many clients, while some of the alternative demonstration methods can require personalization, and as a result, incur high cumulative costs in case many clients request such personalized demonstrations. Depending on the company profile, a cost analysis in such case might favor certification despite its high initial costs. However, if costs pose a barrier to obtaining an official certification, our workshop participants noted that simply following a standard can support the dialogue between them as providers and their clients. While in some cases this may be sufficient, others may require more efforts, so the broader acceptance of this approach needs further investigation. Similarly, costs have been mentioned as an issue for using \textit{tests and reports}, as conducting audits and penetration testing activities can turn out to be expensive, especially if clients ask for updated reports on an annual basis. As pointed out by the participants, and similarly by the finding of Thomas et al. ~\cite{Thomasetal10.1145/3173574.3173836} automation of tools could potentially reduce costs and manual labor spent on testing. However, some of the tools might also require costs in both procurement of the software and the resources required to train developers to use these tools. While monetary costs were not frequently mentioned for conducting \textit{interactive sessions} and answering \textit{questionnaires}, these activities still required resources such as time and human resources. 

\paragraph{Reliability:} 
In terms of accurately representing the security level of a company's products, \textit{certifications} were usually seen as one of the most reliable methods for demonstrating security, even with some participants acknowledging the limitations of their reliability, such as the need to trust the company that it still continuously adheres to the security practices it claimed at the time of getting certified. However, as highlighted in the background section, certifications merely indicate that the certified company meets the minimum criteria established by the certifying body, but they may not independently have the ability to provide comprehensive assurance.~\cite{Guerra2021TheRO, GAITERO2021110960,Chidukwani,cavenetal,chenetal}. Similarly, reports from \emph{audits} performed by reputable companies were seen to reliably demonstrate security of the company and its practices and processes, while at the same time some concerns about the reliability of report’s assurances towards protection against future attacks have been raised. The reliability of the \textit{questionnaires}, on the other hand, depends strongly on the quality of the questionnaire, assuming that the questions are appropriately aligned with the given context. A similar argument can be made for the \textit{interactive sessions}, where the reliability of the results depends on the expertise of the people participating in these sessions and the extent to which the company is ready to demonstrate their internal processes. Finally, \textit{social proof} as a method has a potential of being unreliable, considering its dependence on subjective referrals.

\paragraph{Accessibility to non-experts:} While the content of the standards themselves is often unknown to non-experts, the presence of a \textit{certification} is commonly accepted as a sign of quality, if the certifying body is trusted. Similarly, while the content of \textit{test/audit} reports might be full of technical details, non-experts can rely on the trustworthiness of the organization performing the audits – however, this requires additional knowledge into which audit companies can be considered reputable. As Babar et al. argue~\cite{Baberetal}, factors such as positive references, certifications, and recommendations of reputable agents can be considered vital for the establishment of partnerships. In addition to this, some automated tools feature a color-coded scale (e.g. green/good, red/bad), indicating the results of the latest test, making it easier for receivers with limited technical knowledge to interpret the outcomes. The question of accessibility becomes more challenging, however, when it comes to the \textit{questionnaires} and the \textit{interactive sessions} methods that require an interaction with the client – either before providing a service (questionnaires) or during/after (interactive sessions). As such, while someone without technical expertise can potentially both come up with the questionnaire or participate in an interactive session, their lack of expertise will render the corresponding questionnaire or interactive session unreliable, as mentioned above. Finally, \textit{social proof} can be made accessible to non-experts, while still requiring some work to check the reputation of the company.

These factors show that there is such a variety of methods that companies use, that a single demonstration method is unlikely to work for every company or every stakeholder in need of a demonstration, just as Chidukwani~\cite{Chidukwani} and Guerra and Hinde~\cite{Guerra2021TheRO} advocate that comprehensive and hybrid frameworks would be more effective as compliance certifications alone are insufficient for full assurance. An effective approach towards security demonstration, therefore, would include effectively ways to combining these methods as well as adjusting them to the particular needs of all the involved stakeholders. Furthermore, while we focused on investigating methods for demonstrating security of products and services provided by companies, our interviewees frequently mentioned methods for demonstrating the security of the IT infrastructure of the company and the internal processes in place. While these issues are interconnected, and secure processes are a necessary prerequisite for developing secure software, the optimal choice of a specific demonstration method can differ depending on which of these two aspects needs to be demonstrated.

Finally, our findings indicate that all the demonstration methods described rely on trust, especially if the client or another stakeholder interested in demonstration lacks technical expertise. The role of trust can support the findings of Dalela et al.~\cite{Dalelaetal} which notes how the prevalent trust within companies can mirror the broader importance of trust in society as a whole, but should be further studied in the context of security demonstration methods.

\paragraph{Limitations of our study:} 

In our study we did not aim to evaluate the actual security practices in the companies or to identify internal issues they face, instead focusing on how companies communicate the security they have in place. Nonetheless, our results provide insights into how  existing security measures are shaped by the expectations and requirements of their clients. Although our study only included Danish companies, mainly located in the capital region of Denmark, our participants have strong contacts with international clients and large clients with a variety of providers. Therefore, the challenges pertaining to our participating software development SMEs can be representative of the challenges faced by similar companies in other regions.

Our participants represent mostly micro and small companies (40\% from micro companies, 33,3\% from small companies, and 6,7\% from medium companies), which may not fully capture the complexities encountered by those in the medium enterprise category. Conversely, enterprises outside of this size range, e.g. large enterprises, are likely to face different challenges or adopt other approaches to cybersecurity, given their broader resources and organizational scale. 
 Furthermore, we investigated the question of security demonstrations while looking only at the perspective of the company doing the demonstration. Considering the side of stakeholders in need of demonstration, be it other companies, governmental bodies or individuals using products/services of the company and studying their views on what constitutes a sufficiently reassuring security demonstration, is therefore an important direction of future work.
\section{Recommendations} \label{sec:recommendations}

Drawing from insights gained through the semi-structured interviews and the validation workshop, we derive key recommendations for companies to effectively demonstrate the security measures integrated within their organization and software solutions. The recommendations are organized by demonstration method. We stress that due to a variety of demonstration methods, a dialogue between the SME as service provider and the client is most crucial to ensure a common understanding of which demonstration methods are acceptable for both parties and are the most appropriate choice for a particular use case.

\subsection{Certifications}

As our results demonstrate, certifications are an optimal method of providing security assurance. These certifications are typically recognized either nationally or internationally, meaning that a few select certifications can satisfy the diverse security requirements of multiple clients, who might otherwise make various individual requirements, such as penetration test reports, audit reports, and/or filling out extensive questionnaires. While certifications are costly, for all companies for which this is a barrier, even following the guidelines of a recognized standard without official certification, provides them an acceptable demonstration method for their customers who can recognize standardized principles listed by the certifications. Therefore, depending on the business model, it makes sense for SMEs to consider the cumulative costs and benefits of following the guidelines of a standard, even without an official stamp of certification. However, as with all of the methods, a dialogue with the client is required to understand whether following the guidelines without obtaining a certification is accepted as a sufficient security demonstration.

\subsection{Tests and Reports}

Audit reports and penetration test reports are recognized methods for demonstrating security. Audit reports can be particularly useful in sales related contexts, while penetration test reports are better suited for clients with the technical expertise to interpret them. However, both can be expensive and their limited validity requires regular renewal of these types of reports. Ideally, these reports are complemented by automated tools that provide real-time, comprehensive insights into code security and vulnerabilities. But, for resource-constrained SMEs, adopting affordable or open-source automation tools as an entry point can be a practical approach, and then gradually scaling up to more advanced tools as resources permit to further enhance security demonstrations. Prioritizing automation for addressing the most frequently requested requirements allows SMEs to balance costs and benefits effectively.

\subsection{Questionnaires}
Our study found that some questions are irrelevant or inapplicable, often due to being poorly crafted or copied. We therefore recommend that issuers prioritize transparency and engage with SMEs to clarify questions, tailoring them to avoid overwhelming respondents with an excess of inquiries.
Questionnaires are good at providing insights into an SMEs' processes and security, which is why providing thorough and transparent responses can be particularly advantageous for the supplying SME, as emphasized by our participants as well. Engaging in dialogue to clarify ambiguous questions ensures accuracy, and using feedback as an internal audit can address security gaps. Well-answered questionnaires can also serve as a competitive advantage.

\subsection{Interactive Sessions}
Interactive sessions can be practical alternatives to more rigid or resource-intensive security demonstration techniques, especially for resource constrained SMEs. Offering interactive sessions to clients can enhance transparency, even if they choose not to participate. If the company is reluctant to share some parts of their code and systems with the customer (e.g. due to security concerns), these restrictions should be clearly communicated. The interactive session should involve thorough preparation via scheduling enough time, including relevant experts and providing detailed post-session feedback. The session should be documented for transparency.

\subsection{Social Proof}
Engaging with open-source communities and using widely adopted, reputable components enhances security assurance. Leveraging the reputation of third-party providers and showcasing client testimonials can build trust. However, relying solely on social proof without additional security measure can be counterproductive, that is why social proof should be considered an ideal supplementary method to assess security, but not the primary.

\section{Conclusion} \label{sec:conclusion}

Our work contributes to the growing body of knowledge aimed at supporting SMEs and their ecosystems in effectively communicating about security by systematically examining how SMEs demonstrate the security of their software in a B2B model. The findings, derived from ethnographic interviews and a validation workshop, highlight five key techniques used for security demonstration: Certifications, Tests and Reports, Questionnaires, Interactive Sessions, and Social Proof — each varying in cost, reliability, and accessibility. As each of these techniques comes with its own advantages and disadvantages, a consideration of various factors such as costs or the need for a demonstration to be accessible to non-experts should be considered by both the company and the customer to decide how security should be demonstrated in each separate case.

\paragraph{Future work:} Future research would focus on investigating methods to effectively combine and use the demonstration methods in a way that is both beneficial for the company and the customer. As such, a more thorough studying of the limitations of each demonstration method, including e.g. large-scale and/or longitudinal studies involving diverse stakeholders, will help in identifying and remedying issues that prevent companies from adopting these methods. A further direction of future work would be investigating ways to improve the demonstration methods, such as developing frameworks for more accessible certification processes, standardized questionnaires that would both reduce the workload on behalf of companies filling out the questionnaires and on behalf of the client coming up with the questionnaire, automated test tools with clear outputs for companies with limited resources, or a proposal to conduct interactive sessions in a more structured and rigorous manner. Such improvements would furthermore focus on enhancing the transparency of security demonstration approaches, making them more accessible for non-experts.

\begin{acks}
This research was in part supported by Digital Lead. The opinions expressed and arguments employed herein do not necessarily reflect the official views of the funding body.
\end{acks}




\bibliographystyle{ACM-Reference-Format}
\bibliography{Reference}

\section*{Appendices}

\appendix
\section{Interview guide} \label{appendix:a}

\textbf{Introduction (5 minutes)}

Thank you for taking time out of your day to talk with us. We really
appreciate it. To kick things off, I just want to provide you with
some background on us and why we are doing this interview. I work in a
{Research Center in ITU called Center for Information Security \&
  Trust (CISAT)}.  This is a collaborative research project between
{ITU and our two industry partners; CyberJuice and Security
  Scientist}.  The reason we are doing this research is to better
understand how small and medium size enterprises demonstrate the
security of their software. There is no right or wrong way to answer
the questions. I am just interested in learning about your way of
thinking, so I would like for you to just think aloud as you go along
and express your opinion. Your participation is anonymous. We will not
reveal the names of any participants or companies in our report.

Is it alright if I record the session? The recording will only be shared with the relevant people in [XXX].
The collected data will be deleted after the completion of the project. An anonymized version of the data and quotes will be used for the report and for further publications. 

Any questions before we begin? The interview will take 1-1,5 hours.

The interview is structured in such a way that I will begin by asking about your role in the company, followed by a description of the company and the overall security requirements. Then, the interview is divided into two phases. In the first phase, I will inquire more deeply about security requirements, and in the second phase, we will address communication and demonstration aspects. The interview concludes with any additional remarks you would like to make.\\

\textbf{Opening questions (10 minutes)}

\textit{Opening}

\begin{itemize}
    \item To begin with, can you please start by telling me about yourself and your role in the company?
\end{itemize}

\textit{Present company}

\begin{itemize}
    \item Can you please briefly introduce the company and its products and services?
\end{itemize}
\begin{itemize}
    \item In what markets do you operate? \\ Follow up: Can you describe the profile of your customers, for example the category of customers and the size of their business?
\end{itemize}

\textit{Present requirements}

\begin{itemize}
    \item Could you please tell me if you have any requirements or expectations to demonstrate the security of your mentioned products or services?\\ Follow up: Who imposes such requirements or where do they come from? \\ Follow up: Could you please provide examples of such requirements from your most recent projects where security was relevant?
\end{itemize}

\textbf{Stage 1 (20 minutes)\\}
\textit{Requirements and certifications}

\begin{itemize}
    \item Based on what we just talked about, i.e. security requirements you need to comply to, could you please tell me what your current security practices are?
\end{itemize}
\begin{itemize}
    \item What actions do you take to ensure that the specific security requirements are met for compliance? \\ If “don’t know”: For instance, do you have any checks to ensure that the security is followed? 
Follow up: Do you document them or monitor them? Or are there any other considerations you'd like to make in this regard? 
\end{itemize}
\begin{itemize}
    \item Are there any standards that are particularly relevant for you such as ISO 27001/NIST/other? \\Nudging: You mentioned you do business in the United States for the healthcare sector. Are you aware of any requirements from their side? 
\end{itemize}
\begin{itemize}
    \item Do you have any security requirements that are imposed by more parties? Can you give some examples of who it can be, and do you do anything differently or do you follow different processes to comply with multiple ones?
\end{itemize}
\begin{itemize}
    \item Do you subcontract tasks to (software) third parties? \\ If yes: How do they demonstrate their security to you and how do you communicate this to your customers? \\If not: Do you use any third-party (software) components and how do you ensure its security? 
\end{itemize}
 \begin{itemize}
     \item When developing new product features how do you ensure security in the software development process?
 \end{itemize}
 \begin{itemize}
     \item Do you have any further comments or considerations?\\
 \end{itemize}

\textbf{Stage 2 (20 minutes)\\}
\textit{Communication}

\begin{itemize}
    \item Let’s now focus on how you demonstrate the security of your product/services to relevant parties. First of all, are you required to provide any formal documentation to your clients to show the status of compliance? \\ \\If yes: Can you give examples of types or documentation and the frequency you do this?\\
Follow up: What channels do you typically use?\\
Follow up: Can you give some examples of the most challenging aspects?\\

If not: What else do you do to showcase that your products are secure? \\
Follow up: Can you give examples of some of the most useful aids in this type of communication?\\
\end{itemize}
\begin{itemize}
    \item Can you recall any cases in which changes to security or related concerns  emerged? \\

Follow up: Can you tell me about the process and how you dealt with it at your company?\\
Follow up: What was challenging and what was working well? \\
Follow up: What would you differently with the knowledge or experience that you have now?\\
\end{itemize}

\begin{itemize}
    \item Can you recall cases where you tried to demonstrate security and experienced that it went all smoothly? \\

Follow up: What do you think contributed to the positive outcome? \\
Follow up: Do you think this was product/project specific or can you transfer the knowledge to other domains? \\
\end{itemize}

\begin{itemize}
    \item Can you recall cases where you tried to demonstrate security where it didn't go so well, or it wasn't understood by the recipient? \\

Follow up: Can you tell me what you think went wrong? \\
Follow up: What would you do differently? \\
\end{itemize}

\textbf{Outro}\\
\textit{End of interview}

\begin{itemize}
    \item Do you have any further comments or any other remarks we did not cover that you wish to mention?
\end{itemize}
\end{document}